# SIZE EFFECTS IN FERROELECTRIC THIN FILMS: THE ROLE OF 180° DOMAINS

**Rajeev Ahluwalia**[1] **and David J. Srolovitz**[2]

[1] *Institute of Materials Research and Engineering, Singapore, 117602*

[2] *Department of Physics, Yeshiva University, New York, N.Y. 10033, U.S.A*

Abstract

The depolarization fields set up to due to uncompensated surface charges in a ferroelectric thin film can suppress the ferroelectric phase below a critical size. Recent experiments show that 180° domain structures can help to stabilize ferroelectricity in films as thin as 3 unit cells. We study the influence of domain structures on the size-dependent properties of ferroelectric thin films using time-dependent Ginzburg-Landau theory. The model incorporates the effect of depolarization field by considering non-ferroelectric passive layers at the top and bottom surfaces. We show that the 180° domain size decreases as the film thickness is reduced and the film abruptly becomes paraelectric below a critical size. The 180° domains appear during polarization switching causing a time-dependent relaxation of the remnant polarization $P_r$ , consistent with recent experiments. The depolarization-induced domain wall motion significantly alters the shape of the polarization versus electric field (*P-E*) loops at small thicknesses.

Ferroelectric thin films are becoming increasingly important for applications in memory [1] and piezoelectric devices [2]. As the device size becomes progressively smaller, it becomes important to understand the behavior of ferroelectricity at a nano-scale. Specifically, one needs to understand whether ferroelectricity is still retained when the film thickness is as thin as a few unit cells. At the nano-scale, the bound charges at the electrode-ferroelectric interfaces crucially influence the physics. These charges set up a depolarization field which tends to oppose the polarization. If the surfaces charges are not compensated, this depolarization field can suppress ferroelectricity below a critical size [3]. In thin films, the depolarization effect can be reduced if the surface charge is compensated by the free charge in the conducting electrodes or from the accumulation of charged ions at the free surface [4]. When the surface charge is not compensated, the depolarization fields are reduced by forming equilibrium 180° domains [5-7]. Recent experiments on $PbTiO_3$ films report 180° domains for films as thin as three unit cells [6]. The 180° domain pattern significantly modifies the effective electrical properties of these ultra thin films. Since domain wall motion can also cause relaxation of the remnant polarization $P_r$ during switching [8], it is important to understand the role played by the 180° domains in ultra-thin ferroelectric films.

Recently, Dawber, *et al.* [9] employed scaling analysis to study the effect of depolarization fields on the size dependence of the coercive field in thin films. This analysis, however, did not account for 180° domains. Bratkovsky and Levanyuk [7] investigated 180° domains by considering a passive (non-ferroelectric) layer at the ferroelectric-electrode interface. (Non-ferroelectric layers provide incomplete charge compensation at the interface.) They showed that the passive layer stabilizes a 180° domain pattern instead of a single domain state. Although this calculation showed that the 180° domain patterns could strongly modify the physics in thin film ferroelectrics, their influence on size effects and switching kinetics remains unexplored. Tagantsev, *et al.* studied the effects of passive layers on the switching loops, but here too kinetic and domain wall motion effects were not incorporated [10].

Recent experimental studies of domain patterns [5,6] and polarization switching [8] pose some interesting questions. How does the 180° domain pattern change as the film size is reduced to a just a few nanometers? What is the role of 180° domains during polarization switching and on polarization-electric field (*P-E*) loops? In this Letter, we answer these questions by combining the passive layer model discussed above and a time-dependent Ginzburg-Landau (TDGL) approach that describes kinetic effects. The TDGL equations have been widely used to study domain evolution and switching in ferroelectrics [11-14], but neither in conjunction with passive layers nor 180° domains.

We consider a ferroelectric film of thickness $h$ ($|z| \leq h/2$) with passive (i.e., dielectric) layers of thickness $d$ at the top and bottom surfaces. We assume that the film is c-axis oriented and, hence, the polarization normal to the film surface $P_z$ is the appropriate order parameter. This means that 90° domains cannot form in the present model. The total free energy is given as $F_T = F + F_{dep}$, where

$$F = \begin{cases} \int d\vec{r} \left[ \alpha_0 (T-T_0) P_z^2 - \beta P_z^4 + \gamma P_z^6 + \frac{K}{2} (\nabla P_z)^2 + \frac{1}{2\chi_{xx}} P_x^2 \right] & |z| \leq \frac{h}{2} \\ \int d\vec{r} \left[ \frac{1}{2\chi_p} (P_x^2 + P_z^2) \right] & \frac{h}{2} < |z| \leq \frac{h}{2} + d \end{cases} \quad (1)$$

For $|z| \leq h/2$, the above equation represents a ferroelectric in which $P_z$, the polarization normal to the film surface is the order parameter. A harmonic term in $P_x$, the polarization in the lateral direction is also included, allows for the rotation of the polarization vector in the x-z plane. The quantities $\alpha_0, T_0, \beta, \gamma$ and $K$ are material specific parameters that describe the ferroelectric transition. $\chi_{xx}$ is the susceptibility in the direction parallel to the film surface, $x$. For $h/2 < |z| \leq (h/2+d)$ (inside the passive layers), the free energy is that of an isotropic linear dielectric with susceptibility $\chi_p$. The depolarization energy associated with the film is given by

$$F_{dep} = -\int d\vec{r} \left[ \frac{\varepsilon_0}{2} (\vec{E}_{dep} \bullet \vec{E}_{dep}) + \vec{E}_{dep} \bullet \vec{P} \right], \quad (2)$$

where the depolarization field is related to the potential $\phi$ by $E_{dep} = -\vec{\nabla}\phi$. The boundary conditions for the film are $\partial P_z / \partial z = 0$ at $|z| = h/2 + d$. The potential is specified at the surfaces as $\phi = -V/2$ at $z = h/2 + d$ and $\phi = V/2$ at $z = -(h/2+d)$, where $V$ is the applied voltage across the film. Further, the Maxwell equation $\vec{\nabla} \bullet \vec{D} = 0$ ($\vec{D} = \varepsilon_0 \vec{E} + \vec{P}$) gives rise to the constraint

$$-\varepsilon_0 \nabla^2 \phi + \vec{\nabla} \cdot \vec{P} = 0 \ . \tag{3}$$

The TDGL equations for the evolution of the polarization are given as

$$\frac{\partial P_i}{\partial t} = -\Gamma \frac{\delta F_T}{\delta P_i} \qquad i = x, z \ . \tag{4}$$

$\Gamma$ is related to the domain wall mobility and sets the simulation time scale.

Equations (1)-(4) are used to study the domain structures and switching process in ferroelectric films. The parameters were chosen to be appropriate for a BaTiO$_3$ single crystal [14], $\alpha_0$=3.34x10$^5$ Vm/C, $T_0$=381K, $\beta$=6.381x10$^8$ Vm$^5$C$^{-3}$, $\gamma$=7.89x10$^9$Vm$^9$C$^{-5}$, and $K$=1.38x10$^{-11}$Vm$^3$C$^{-1}$, and the temperature was set to $T$=300K. The susceptibilities are chosen as $\chi_{xx}$=417$\varepsilon_0$ and $\chi_p$=208$\varepsilon_0$ for illustrative purposes. The simulations are performed in 2-D using finite difference methods with a simulation cell that is periodic along the $x$-direction and the condition $\partial P_z / \partial z = 0$ at $z = \pm(h/2 + d)$. The smallest length scale in the simulations is $\delta$=0.71nm and the lateral width is $L$=177.5nm. The thickness of the passive layer is fixed at $d$=7.1nm while the film thickness is varied in the range $7.1nm \leq h \leq 60.3nm$.

We first study the formation of domain structures following a quench from the paraelectric phase, as a function of the film thickness $h$. Time is measured in the rescaled units, $t^* = (81\alpha_0 \Gamma)t$. The paraelectric state is simulated by initializing $P_x(\vec{r}, t=0)$, $P_y(\vec{r}, t=0)$ and $\phi(r, t=0)$ with small amplitude random fluctuations around zero. We use short circuit boundary conditions, i.e., $\phi(x,z=h/2+d) = \phi(x,z=-(h/2+d))=0$. With these boundary conditions, Eqs. (3)-(4) are used to simulate the formation and the evolution of the domain structures. Figure 1 shows the simulated 180° domain patterns in films of thickness $h = 60.35nm$, $h = 24.85nm$ and $h = 7.1nm$. The domain width decreases with the decreasing film thickness and the film becomes paraelectric for thicknesses below $h = 7.1nm$. Figure 1(c) shows the thinnest film ($h = 7.1nm$) that still exhibits ferroelectric domains. Note that the domain wall thickness for this case appears to be larger than in the thicker films in Figs. 1(a) and 1(b), indicating the proximity to a size-induced ferroelectric to paraelectric transition.

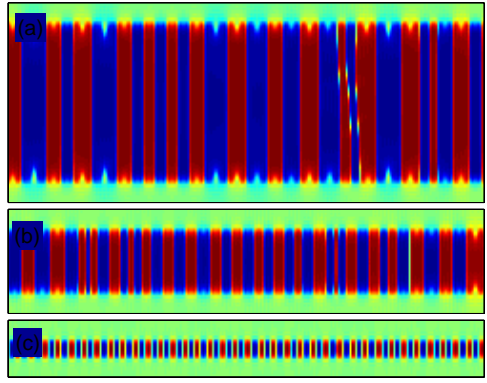

**Fig 1**: (color online) Domain patterns for films with thickness (a) $h = 60.35nm$, (b) $h = 24.85nm$ and (c) $h = 7.1nm$. The passive layer width $d = 7.1nm$ and the lateral size of the simulation cell $L = 177.85nm$ are kept fixed. The polarization in the domains varies between $P_z = 0.26 Cm^{-2}$ (red) and $P_z = -0.26 Cm^{-2}$ (blue). Note that the domain width for the $h = 7.1nm$ film is still much larger than the grid spacing.

The distribution of the depolarization field $E_z = -(\partial\phi/\partial z)$ for the domain patterns in Fig.1 is shown in Fig 2. Note that the depolarization field is largely localized at the passive layer-ferroelectric interface. This figure clearly shows that domain formation shields the bulk of the film from the depolarization field. For the $h = 7.1nm$ film, the field from the top interface nearly overlaps that from the bottom interface. This already corresponds to a high electrostatic energy and, hence, below this size, the film becomes paraelectric.

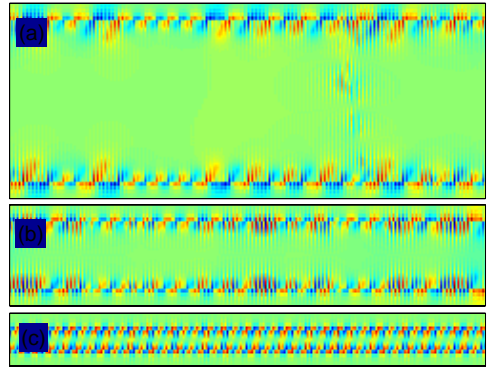

**Fig 2**: (color online) The distribution of the depolarization field $E_z$ corresponding to Fig. 1. $E_z$ varies between (a) $\pm 1200 kV/cm$, (b) $\pm 1100 kV/cm$ and (c) $\pm 590 V/cm$.

Next we study the effect of changing the film thickness on the polarization switching process. At first, the method used to obtain the domain structures of Fig. 1 is repeated but this time with an applied voltage $\phi(x,z=h/2+d) = -V_m/2$ and $\phi(x,z=-(h/2+d)) = V_m/2$. We will refer to this process as field cooling. The configuration formed after this process is then subjected to a time-dependent potential to simulate the switching process. The switching field is applied as $\phi(x,z=h/2+d) = -(V_m/2)\cos(\omega^* t^*)$ and $\phi(x,z=-(h/2+d)) = (V_m/2)\cos(\omega^* t^*)$, where $\omega^*$ is the switching frequency. The $P$-$E$ loops are calculated by plotting the average electric field $<E_z>$ versus the average polarization $<P_z>$, where the averages are performed over the entire dielectric-ferroelectric-dielectric composite.

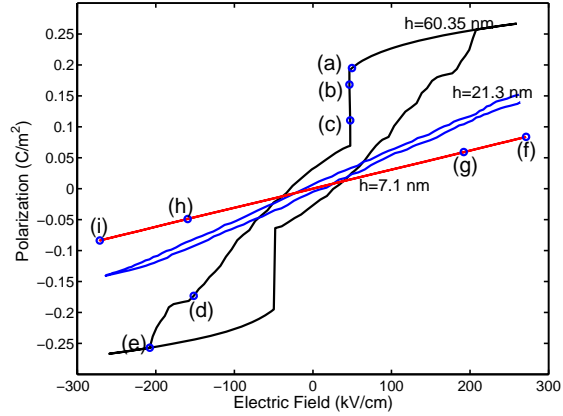

**Fig. 3**: (color online) Simulated $P$-$E$ curves at frequency $\omega^* = 3.14 \times 10^{-6}$ for $h = 60.35nm$, $h = 21.3nm$ and $h = 7.1nm$. The domain patterns corresponding to points (b)-(d) are shown in Fig. 4 and for (f)-(i) are shown in Fig. 5. The point (a) for $h = 60.35nm$ corresponds to the unswitched single domain state with $P_z > 0$ and (e) corresponds to a fully switched single domain with $P_z < 0$.

Figure 3 shows the calculated $P$-$E$ curves for films with thickness $h = 60.35nm$, $L = 21.3nm$ and $h = 7.1nm$ at a switching frequency $\omega^* = 3.14 \times 10^{-6}$ and potential $V_m$ chosen such that $<E_z^{max}> \sim 260 kV/cm$ for all thicknesses. Note, the large $P$-$E$ loop shape change with changes in the ferroelectric thickness. This is consistent with recent experiments on $BaTiO_3$ films [15] and can be understood by examining the domain evolution shown in Figs. 4 and 5. The domain patterns corresponding to points (b)-(d) are shown in Fig. 4 for $h = 60.35nm$ ((a) and (e) correspond to single domain states). For this case, the field cooling at $<E_z^{max}> = 260kV/cm$ results in a single domain state with $P_z > 0$. As the electric field is reversed, there is a depolarization induced single- / multi-domain transition at a critical field which leads to a jump in the *P-E* curve. The transition occurs by formation of the reversed polarization domains in the middle of the film. Thereafter, as the field is further decreased, the reversed domains grow by sideways motion and eventually a single domain of the reversed polarization is obtained. This depolarization induced domain formation and domain wall motion is responsible for the unusual shape of the *P-E* curve. Since we focus on depolarization effects, we ignore the nucleation mechanisms associated with thermal noise and defects. In a real system, thermal noise and defects will cause domain nucleation and the abrupt transition observed in the present simulations may not show up as clearly in the *P-E* loops. We also note that as we increase the film thickness, the depolarization effect weakens and we obtain "squarer" loops at larger sizes.

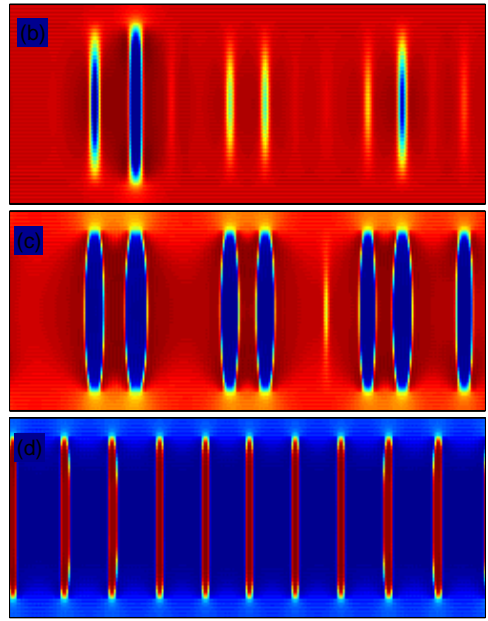

**Fig 4**: (color online) Domain patterns for $h = 60.35nm$, corresponding to points (b), (c) and (d) in Fig. 3.

At small thicknesses ($h < 24.85\ nm$), the depolarization effect is so strong that for the same applied external field $<E_z^{max}> = 260 kV/cm$, 180° domains are observed instead of a single domain after field cooling. In these cases, domains are present during the entire switching process. We see in Fig. 3 that $P_r$ decreases with decreasing ferroelectric thickness, eventually leading to a linear *P-E* curve without any hysteresis for $h = 7.1 nm$. In this case, the switching process may be described as the back and forth sideways motion of the domain walls, as seen in Fig. 5.

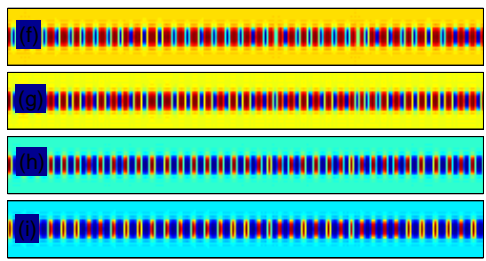

**Fig 5:** (color online) Domain patterns for $h = 7.1 nm$, corresponding to points (f), (g), (h) and (i) in Fig. 3.

Figure 6 shows the simulated hysteresis loops for a case with a particularly large passive layer susceptibility, $\chi_p = 2080\varepsilon_0$ (all other parameters as in the previous simulations) for a film with thickness $h = 7.1 nm$. The two loops correspond to different switching frequencies. A comparison of the *P-E* loops with that in Fig. 3 for $h = 7.1 nm$ shows that the hysteresis is larger for the case with higher passive layer susceptibility. This can be attributed to the smaller polarization gradients at the passive layer – ferroelectric interface that result in a smaller depolarization field in the larger passive layer susceptibility case. This suggests that high susceptibility dielectric layers may help stabilize ferroelectric behavior to very small ferroelectric layer thicknesses. Figure 6 also suggests that $P_r$ and the hysteresis loop shape depend upon switching frequency. For slow switching, there is sufficient time for the depolarization-induced domain formation to set in, leading to a reduced $P_r$. These results are in good qualitatively agreement with the recent experimental observations by Kim, *et al.*, where a similar frequency dependence of $P_r$ was observed for thin $BaTiO_3$ films [8]. We note that the previous analytical approaches [7,10] are not capable of capturing such important kinetic effects.

In summary, we have studied ferroelectric film thickness effects within the framework of a TDGL model, incorporating both the effects of depolarization fields and 180˚ domains. We demonstrated that depolarization-induced 180˚ domain formation strongly modifies the ferroelectric behavior of nano-scale films. The motion of domain walls results in a significant shape change of the *P-E* loops as the film thickness is reduced, consistent with recent experiments [15]. The present calculation also underscores the importance of using a kinetic approach to understand the physics of ferroelectric thin films, particularly in light of recent experiments [8] that show that time dependent effects become very important at the nanoscale.

We gratefully acknowledge useful discussions with Nikolai Yakovlev and the A-Star VIP programme for financial support.

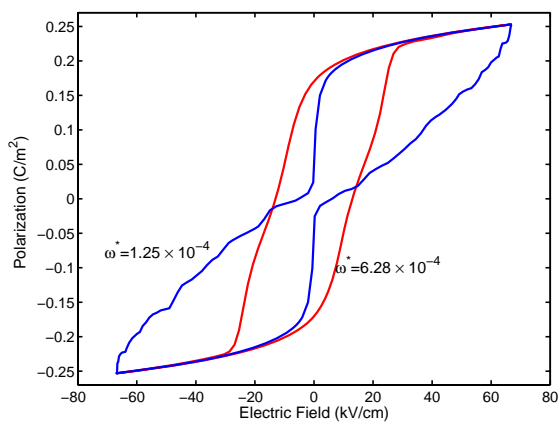

**Fig. 6**: (color online) Simulated *P-E* loops for the case with $\chi_p = 2080\varepsilon_0$ and $h = 7.1 nm$ at two different switching frequencies.